\documentclass[letter]{jpsj2}

\def\runtitle{
Quartet Fermi Superfluidity
}
\def\runauthor{Kazumasa {\sc Miyake}}

\title{
On Quartet Superfluidity of Fermionic Atomic Gas
}          
\author{Hidekazu {\sc Kamei} and Kazumasa {\sc Miyake}}
\inst
{
Division of Materials Physics, Department of Materials Engineering Science \\
Graduate School of Engineering Science, Osaka University, Toyonaka, 
Osaka 560-8531
}

\recdate
{April 7, 2005}

\abst
{
The possibility of a quartet superfluidity in fermionic systems is studied 
as a new aspect of atomic gas at ultralow temperatures.  
The fourfold degeneracy of a hyperfine state and moderate coupling 
is indispensable for the quartet superfluidity to occur.  Possible 
superconductivity with quartet condensation in electron systems is 
discussed.  
}

\kword{
quartet Fermi superfluidity, laser-cooled fermionic atomic gas, 
plasmon mechanism superconductivity
}
\begin{document}
\sloppy
\maketitle
The superfluidity is so far known to be sustained by the Bose-Einstein 
condensate or the Cooper-pair condensate.  The former is realized in liquid 
$^4$He and some atomic gas of an alkali element~\cite{BEC1,BEC2,BEC3,BEC4}, 
while the latter is realized in liquid $^3$He 
and a variety of superconductors.  Recently, it has been found that 
Cooper pairs are formed in the fermionic atomic gas of alkali 
element~\cite{Regal} and the crossover to the Bose-Einstein condensation of 
a diatomic molecule is possible with the help of Feshbach 
resonance~\cite{Ohashi}.  

In principle, there exists the other possibility that superfluidity is 
sustained by a condensate based on four fermions (quartet) as in the 
$\alpha$-particle correlation in a light nucleus~\cite{Tohsaki}.  
An $\alpha$-particle consists of two protons and 
two neutrons which have approximately 4-fold degeneracy corresponding to 
$2\times2$ degeneracy of the real spin and isotopic spin states.  
Such 4-fold degeneracy is possible in a fermionic atomic gas with total 
angular momentum $F=S+I=3/2$, $S$ and $I$ being the electron spin and 
nuclear spin, respectively.  For example, a Be atom has $F=3/2$ with $S=0$ 
and $I=3/2$.  

The purpose of this Letter is to assess the possibility of 
quartet superfuidity by analyzing an extended Cooper 
problem~\cite{Cooper}, i.e., a four-fermion problem outside a {\it rigid} 
Fermi surface.  This is regarded as a natural extension of the Cooper problem 
to the present case.  We expect that the same tendency as in the 
``Cooper problem" is taken over to the many-body problem in which the 
deformation of the Fermi surface should be taken into account 
self-consistently.  One might think that 
there is another way of extending the Cooper problem to the present case, 
i.e., by analyzing a stability condition of the BCS state against an 
addition of four extra fermions to the BCS state.  However, we do not take 
such an approach here because not only mathematical treatments but also 
the concept of a theoretical framework seems to be much more complicated 
than ours.  The result of this study might also give us a clue to 
the possibility of quartet superfluidity in the electronic system with a 
nearly degenerate Fermi surface.  

A theorem by Nagaoka and Usui~\cite{Nagaoka} states that 
the ground-state wave function of a many-particle system is symmetric 
with respect to an interchange of position between any pair of particles 
if one could neglect the possible spin dependence.  Therefore, the ground 
state of four fermions with ``spin" $F=3/2$ should be 
\begin{equation}
\Psi_{\rm Q}(1,2,3,4)=
\Phi({\vec r}_{1},{\vec r}_{2},{\vec r}_{3},{\vec r}_{4})
\chi(1,2,3,4),
\label{eq:1}
\end{equation}
where $\Phi$ is a fully symmetric function with respect to the interchange of 
${\vec r}_{i}\leftrightarrow{\vec r}_{j}$, and the wave function for the spin 
is given by the fully antisymmetric one: 
\begin{equation}
\chi(1,2,3,4)=
{1\over \sqrt{4!}}
\left|
\begin{array}{cccc}
\alpha(1) & \beta(1) & \gamma(1) & \delta(1) \\
\alpha(2) & \beta(2) & \gamma(2) & \delta(2) \\
\alpha(3) & \beta(3) & \gamma(3) & \delta(3) \\
\alpha(4) & \beta(4) & \gamma(4) & \delta(4)
\end{array}
\right|.
\label{eq:2}
\end{equation}
Here, $\alpha$, $\beta$, $\gamma$, and $\delta$ denote the ``spin" states 
$F_{z}=3/2$, $1/2$, $-1/2$, and $-3/2$, respectively.  
Due to the Pauli principle, 
the orbital part $\Phi({\vec r}_{1},{\vec r}_{2},{\vec r}_{3},{\vec r}_{4})$ 
is a fully symmetric function of ${\vec r}_{i}$.  State (\ref{eq:1}) 
represents that of a tetra-atomic molecule (quartet) if the bound-state 
solution is possible, and is more stable than 
that of two diatomic molecules whose orbital wave function is not fully 
symmetric.  

Now, we discuss a four-body problem outside the rigid Fermi surface, 
namely, the Cooper problem.  Here, we compare the energy of the 
quartet and two Cooper pairs.  Because it is difficult in general to obtain 
the exact solution of a four-body problem, we try to solve the problem 
variationally.  The variational solution for $\Phi$ is assumed to be in 
the Hartree form as 
\begin{equation}
\Phi({\vec r}_{1},{\vec r}_{2},{\vec r}_{3},{\vec r}_{4})
=\prod_{i=1}^{4}f({\vec r}_{i}-{\vec R}),
\label{eq:3}
\end{equation}
where ${\vec R}\equiv\sum_{i=1}^{4}{\vec r}_{i}/4$ is the center of 
gravity.  The single-particle state $f({\vec r})$ is expressed as 
\begin{equation}
f({\vec r})=\sum_{|{\vec k}|>k_{\rm F}}f_{\vec k}\,
e^{{\rm i}{\vec k}\cdot{\vec r}}.
\label{eq:4}
\end{equation}
It is noted here that only the states outside the Fermi surface 
$|{\vec k}|>k_{\rm F}$ are available.  

The energy for state (\ref{eq:3}) is given as 
\begin{eqnarray}
E_{\rm Q}&=&\int\cdots\int{\rm d}{\vec r}_{1}\cdots{\rm d}{\vec r}_{4}
\prod_{i=1}^{4}f({\vec r}_{i}-{\vec R})\left[
-\sum_{i=1}^{4}{\hbar^{2}\nabla^{2}\over 2m}+\sum_{i>j}
V({\vec r}_{i}-{\vec r}_{j})\right]\prod_{i=1}^{4}f({\vec r}_{i}-{\vec R})
\nonumber
\\
&=&4\int{\rm d}{\vec r}f({\vec r})\left[-{\hbar^{2}\nabla^{2}\over 2m}
+{3\over 2}\int{\rm d}{\vec r}^{\prime}V({\vec r}-{\vec r}^{\prime})
|f({\vec r}^{\prime})|^{2}\right]f({\vec r}),
\label{eq:5}
\end{eqnarray}
where $V({\vec r}_{i}-{\vec r}_{j})$ denotes the two-body interaction 
between fermions.  
Optimization of $E_{\rm Q}$ with respect to $f({\vec r})$ 
leads to the equation
\begin{equation}
\left[-{\hbar^{2}\nabla^{2}\over 2m}
+3\int{\rm d}{\vec r}^{\prime}V({\vec r}-{\vec r}^{\prime})
|f({\vec r}^{\prime})|^{2}\right]
f({\vec r})=\lambda f({\vec r}),
\label{eq:6}
\end{equation}
where $\lambda$ is the eigenvalue.  This equation should be solved 
under the normalization constraint
\begin{equation}
\int{\rm d}{\vec r}[f({\vec r})]^{2}=1.
\label{eq:7}
\end{equation}
The resultant expression of optimized energy is given by 
\begin{equation}
(E_{\rm Q})_{\rm min}=2\left[\lambda-
\int{\rm d}{\vec r}f({\vec r})\left(-{\hbar^{2}\nabla^{2}\over 2m}\right)
f({\vec r})\right]
+4\int{\rm d}{\vec r}f({\vec r})\left(-{\hbar^{2}\nabla^{2}\over 2m}\right)
f({\vec r}).
\label{eq:8}
\end{equation}

In order to discuss the ``Cooper-type problem", let us introduce the 
$q$-representation for the interactions as
\begin{equation}
V({\vec r})=\sum_{\vec k}V_{\vec q}e^{{\rm i}{\vec q}\cdot{\vec r}},
\label{eq:9} 
\end{equation}
and
\begin{equation}
U({\vec r})\equiv
3\int{\rm d}{\vec r}^{\prime}V({\vec r}-{\vec r}^{\prime})
|f({\vec r}^{\prime})|^{2}
=\sum_{{\vec k}}U_{\vec q}e^{{\rm i}{\vec q}\cdot{\vec r}},
\label{eq:10}
\end{equation}
where $U_{\vec q}$ is expressed in terms of $V_{\vec q}$ and $f_{\vec k}$ 
as
\begin{equation}
U_{\vec q}=3V_{\vec q}\sum_{{\vec p}}f_{{\vec q}-{\vec p}}f_{\vec p}.
\label{eq:11}
\end{equation}
Then, the effective Schr\"odinger equation for the `Cooper problem" is 
given as 
\begin{equation}
\epsilon_{k}f_{\vec k}+\sum_{{\vec k}^{\prime}}
U_{{\vec k}-{\vec k}^{\prime}}f_{{\vec k}^{\prime}}
=\lambda f_{\vec k},
\label{eq:12}
\end{equation}
where $\epsilon\equiv\hbar^{2}k^{2}/2m$.  It is noted that the wave vector in 
the wave function $f_{\vec k}$ is restricted in a way such that 
$k_{\rm F}<|{\vec k}|$, $k_{\rm F}$ being the Fermi wave number.  

In order to solve eq. (\ref{eq:12}), following BCS, we introduce a model 
attractive interaction as follows: 

\begin{eqnarray}
V_{{\bf k},{\bf k}'}&=& 
\left\{
\begin{array}{@{\,}ll}
-V, &(\epsilon_{\rm F}<\epsilon_{k},\epsilon_{k'}<\epsilon_{\rm c});\\
0, &({\rm otherwise}).
\end{array}
\right.
\label{eq:13}
\end{eqnarray}

Then, the second term of eq. (\ref{eq:12}) is expressed as 
\begin{eqnarray}
\sum_{{\vec k}^{\prime}}
U_{{\vec k}-{\vec k}^{\prime}}f_{{\vec k}^{\prime}}&=&
{1\over 4\pi^{2}}\int_{k_{\rm F}}^{k_{\rm c}}{\rm d}k'k^{'2}f(k')
\int_{-1}^{1}{\rm d}tU(|{\vec k}-{\vec k}'|)
\nonumber
\\
&\equiv&
\int_{k_{\rm F}}^{k_{\rm c}}{\rm d}k'U^{*}(k,k')k^{'2}f(k'),
\label{eq:14}
\end{eqnarray}
where $|{\vec k}-{\vec k}'|=(k^{2}+k^{'2}-2kk't)^{1/2}$, i.e., 
$t\equiv ({\vec k}\cdot{\vec k}')/kk'$, and 
the kernel $U^{*}$ is defined as 
\begin{equation}
U^{*}(k,k')={1\over 4\pi^{2}}\int_{-1}^{1}{\rm d}tU(|{\vec k}-{\vec k}'|).
\label{eq:15}
\end{equation}
Here, the function $U(|{\vec k}-{\vec k}'|)$, (\ref{eq:11}), is expressed 
as 
\begin{equation}
U(|{\vec k}-{\vec k}'|)=3V_{{\vec k}-{\vec k}'}{1\over 4\pi^{2}}
\int_{-1}^{1}{\rm d}t'\int_{k_{\rm F}}^{k_{\rm c}}{\rm d}pp^{2}
f(p)f(|{\vec k}-{\vec k}'-{\vec p}|),
\label{eq:16}
\end{equation}
where $t'\equiv ({\vec p}\cdot({\vec k}-{\vec k}'))/p|{\vec k}-{\vec k}'|$.  

Finally, by introducing the function $F(k)\equiv kf(k)$, eigenvalue equation 
(\ref{eq:12}) is transformed to the following form:  
\begin{equation}
\epsilon_{k}F(k)+\int_{k_{\rm F}}^{k_{\rm c}}{\rm d}k'
kU^{*}(k,k')k'F(k')=\lambda F(k).
\label{eq:17}
\end{equation}
Now, the kernel $kU^{*}(k,k')k'$ is symmetric with respect to the 
interchange $k\leftrightarrow k'$.  Then, the eigenvalue problem of 
eq. (\ref{eq:17}) is easily solved numerically if the "interaction kernel" 
$U^{*}(k,k')$ is given.  We solve numerically the self-consistent set of 
equations, (\ref{eq:15})$\sim$(\ref{eq:17}).  Then, the energy of the 
quartet (eq. (\ref{eq:8})) is given as 
\begin{equation}
(E_{\rm Q})_{\rm min}=2\left(\lambda+\int_{k_{\rm F}}^{k_{\rm c}}
{{\rm d}k\over 2\pi^{2}}\epsilon_{k}|F(k)|^{2}\right).
\label{eq:18}
\end{equation}

\begin{figure}[h]
\begin{center}
\rotatebox{0}{\includegraphics[width=1.0\linewidth]{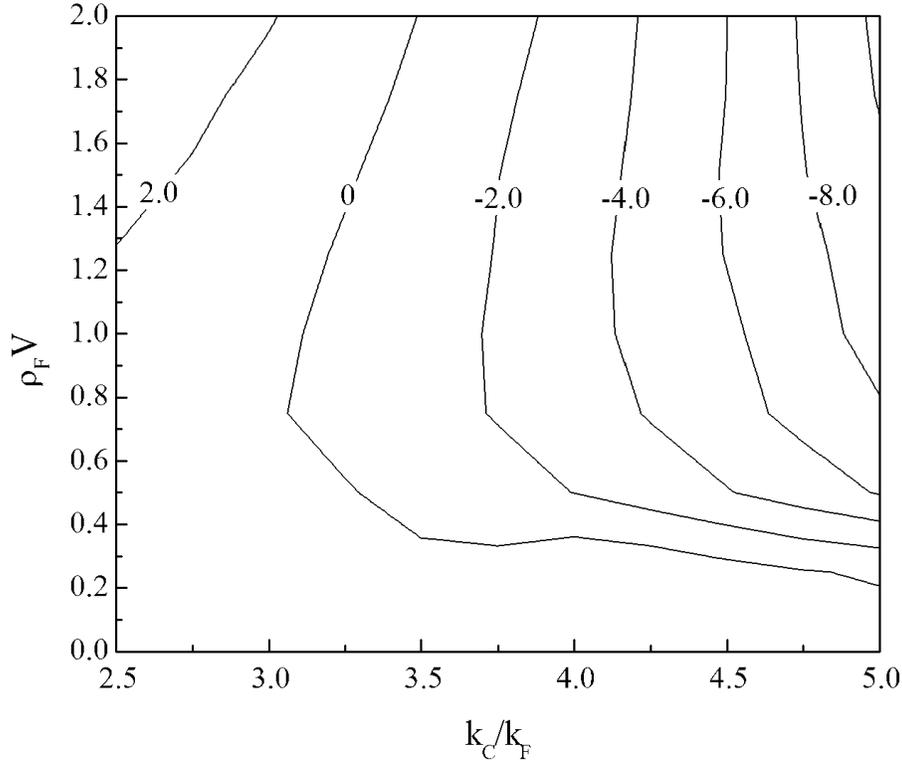}}
\caption{
Phase diagram in $\rho_{\rm F}V$-($k_{\rm c}/k_{\rm F}$) plane.  Numbers 
attached to the lines denote the energy difference ($E_{\rm Q}-2E_{\rm C}$) 
in the unit of $\epsilon_{\rm F}$S.  
}
\label{Fig:1}
\end{center}
\end{figure}

This energy should be compared with that in the original Cooper problem.  
The energy of one Cooper pair $E_{\rm C}$ is given by 
\begin{equation}
1={V\over 4\pi^{2}}\int_{k_{\rm F}}^{k_{\rm c}}{\rm d}k
{k^{2}\over \epsilon_{k}-E_{\rm C}/2}.  
\label{eq:19}
\end{equation}
The phase diagram in the $\rho_{\rm F}V$-($k_{\rm c}/k_{\rm F}$) plane, 
$\rho_{\rm F}\equiv mk_{\rm F}/2\pi^{2}\hbar^{2}$, is 
given in Fig.\ \ref{Fig:1} in which equi-energy surface of 
$(E_{\rm Q}-2E_{\rm C})/\epsilon_{\rm F}$ is shown.  It is seen that 
the quartet state is stabilized in the intermediate and strong coupling 
region with $k_{\rm c}>3k_{\rm F}$.  This is consistent with the Nagaoka-Usui 
theorem which indicates the quartet state to be more stable than the 
two-Cooper-pair state in the dilute limit, i.e., without the Fermi see.  
It is noted that there exists a threshold for eq. (\ref{eq:17}) to have a 
bound-state solution, while there exists no threshold for the original 
Cooper problem (eq. (\ref{eq:19})).  Therefore, in the limit of weak coupling, 
the fully symmetric solution in the orbital space will not be the ground 
state of the ``Cooper problem", in contrast to the 4-body problem in vacuum.  
It is noted that the quantitativeness of the phase boundary in 
Fig.\ \ref{Fig:1} may not be very good because we adopted the Hartree 
approximation to solve the 4-body problem and neglected the 
interaction between two Cooper pairs as in the BCS approximation.  
Nevertheless, this gives us a guideline to understand the crossover between 
the Cooper-pair and the quartet condensations.  

Although we adopted here the model interaction (\ref{eq:13}), it is not so 
difficult to discuss on the basis of a more realistic interaction similar 
to the 12-6 Lennard-Jones potential, as we will discuss elsewhere.  
Since the diatomic bound state is barely possible for the $^4$He 
system~\cite{Bruch}, the weakest interaction and lightest mass, 
it is possible for a neutral atom with a stronger attraction and a 
heavier mass, such as Be atom, to form the diatomic bound state.  
Then, due to the Nagaoka-Usui theorem, the quartet bound state is also 
expected to be possible.  In addition, the tuning of attractive interaction may be 
possible with the use of a mechanism similar to the Feshbach resonance 
for two-body interaction~\cite{Loftus,Roberts,Tiesinga}.  This would make 
it possible to search a crossover phenomenon between the Bose-Einstein 
condensation of tetra-atomic molecules and the quartet Fermi superfluidity.  

In order to discuss the relative stability between the Cooper-pair 
and quartet condensed state, we of course must develop a 
self-consistent treatment that relaxes the rigidity of the Fermi surface.  
Let us briefly sketch how to describe the many-body quartet condensed state.  
A creation operator of a 4-atom molecule with zero total momentum is given as 
\begin{equation}
b^{\dagger}\equiv\sum_{k_{1}}\cdots\sum_{k_{4}}
\delta({\vec k}_{1}+{\vec k}_{2}+{\vec k}_{3}+{\vec k}_{4})
\Phi({\vec k}_{1},{\vec k}_{2},{\vec k}_{3},{\vec k}_{4})
a_{{\vec k}_{1}\alpha}^{\dagger}a_{{\vec k}_{2}\beta}^{\dagger}
a_{{\vec k}_{3}\gamma}^{\dagger}a_{{\vec k}_{4}\delta}^{\dagger}.
\end{equation}
It is natural to assume that the wave function $\Phi$ is given by a 
product of a one-particle state as in eq. (\ref{eq:3}): 
\begin{equation}
\Phi({\vec k}_{1},{\vec k}_{2},{\vec k}_{3},{\vec k}_{4})=
\prod_{i=1}^{4}\phi({\vec k}_{i}),
\label{eq:20}
\end{equation}
Then the many-body quartet condensed state $|\Psi_{\rm QC}\rangle$ may be 
written in a manner analogous to the BCS state as 
\begin{equation}
|\Psi_{\rm QC}\rangle={\cal P}_{N/4}
\exp\left(\sqrt{{N\over 4}}b^{\dagger}\right)|{\rm vac}\rangle,
\label{eq:21}
\end{equation}
where $N$ is the total number of fermions, the projection operator 
${\cal P}_{N/4}$ 
works to project out the $N/4$-quartet state, and 
$|{\rm vac}\rangle$ denotes the vacuum.  State (\ref{eq:21}) is equivalent to 
\begin{equation}
|\Psi_{\rm QC}\rangle={\cal P}_{N/4}\prod_{\{{\vec k}_{1}\sim{\vec k}_{4}\}}\left(\prod_{i=1}^{4}u({\vec k}_{i})
+ \prod_{i=1}^{4}v({\vec k}_{i})\times
a_{{\vec k}_{1}\alpha}^{\dagger}a_{{\vec k}_{2}\beta}^{\dagger}
a_{{\vec k}_{3}\gamma}^{\dagger}a_{{\vec k}_{4}\delta}^{\dagger}\right)
|{\rm vac}\rangle,
\label{eq:22}
\end{equation}
where the factor $\delta({\vec k}_{1}+{\vec k}_{2}+{\vec k}_{3}+{\vec k}_{4})$ 
has been abbreviated in the product $\prod_{i=1}^{4}$, and 
$\phi({\vec k})=v({\vec k})/u({\vec k})$ and the normalization condition 
$|u({\vec k})|^{2}+|v({\vec k})|^{2}=1$ should be fulfilled.  The variational 
functions $u({\vec k})$ and $v({\vec k})$ should be determined so as to 
minimize the free energy, as in the BCS theory~\cite{BCS}.  

Finally, let us discuss the possibility of observing quartet 
superconductivity in solids.  The most promising case is in metals 
with two almost degenerate bands, e.g., the $a$- and $b$-bands, 
and an extremely low carrier density.  In such a system, quasi-4-fold 
degeneracy is maintained for each ${\bf k}$-point by spin $\sigma=\pm$ 
and band index $m=a, b$.  Diluteness of carriers makes it possible 
for the plasmon mechanism to work~\cite{Takada,Kohno}, and the energy 
range of attraction 
$\hbar\omega_{\rm pl}$ extends over the Fermi energy $\epsilon_{\rm F}$: 
\begin{equation}
{\hbar\omega_{\rm pl}\over\epsilon_{\rm F}}={4\over \sqrt{3}\pi}{r_{0}
\over a_{\rm B}}>>1, 
\label{eq:25}
\end{equation}
where $r_{0}$ and $a_{\rm B}$ denote the mean distance between electrons and 
the Bohr radius, respectively.  

In conclusion, the possibility of quartet Fermi superfluidity and 
superconductivity have been investigated on a model interaction of 
BCS type.

\section*{Acknowledgements}
One of us (K.M.) has benefited from stimulating conversations with 
Hiroshi Fukuyama.  
This work was supported by a Grant-in-Aid for Creative Scientific Research 
(15GS0213), and a Grant-in-Aid for Scientific 
Research (No. 16340103) and the 21st Century COE Program (G18) 
by Japan Society for the Promotion of Science.

\vskip24pt
\noindent
{\bf Note added:}

After the present paper had been accepted for publication, we became aware of 
the previous works which discussed the physics similar to 
the present one.  Transition temperature of the quartet condesation 
in nuclear matter has been discussed 
in G. R\"opke, A. Schnell, P. Schuck and P. Nozieres: 
Phys. Rev. Lett. {\bf 80} (1998) 3177.  Model system 
in the large spin limit has been analyzed 
in A. S. Stepanenko and J. M. F. Gunn: cond-mat/9901317.  
The problem of one-dimensional models has also been discussed in 
P. Schlottmann: J. Phys.: Condens. Matter {\bf 6} (1994) 1359; and 
C. J. Wu: cond-mat/0409247.  We thank Congjun Wu who directed our attention 
to those works.  
\end{document}